\documentclass[11pt]{article}
\usepackage{blois,epsfig}

\def\be{\begin{equation}}
\def\ee{\end{equation}}
\def\bea{\begin{eqnarray}}
\def\eea{\end{eqnarray}}

\newcommand{\del}{\partial}

\begin{document}
\vspace*{4cm}
\title{Chiral anomalies in superfluid hydrodynamics}

\author{Yasha Neiman, Yaron Oz}

\address{Raymond and Beverly Sackler School of Physics and Astronomy, Tel-Aviv University, Tel-Aviv 69978, Israel}

\maketitle\abstracts{
This is a brief report of work performed in arXiv:1106.3576. We consider the chiral transport terms in a relativistic charged superfluid, and their relation to triangle anomalies. The terms allowed by the Second Law of thermodynamics have been worked out. A simplified form is proposed on heuristic grounds, from an analysis of the better-understood chiral effects in normal fluids. We point out the appearance of a ``chiral electric conductivity'', and relate it through an educated guess to the axial anomaly coefficient.}

\section{Introduction}

There is currently a revival of interest in hydrodynamics within the high-energy physics community. In particular, relativistic fluids governed by nuclear forces open up possibilities for new dynamical effects. Experimentally, the drive to understand these effects is currently motivated by the study of the quark-gluon plasma produced in RHIC and in ALICE. Other conceivable applications are to neutron-star matter,\cite{NStars} the hot and early universe, as well as yet-to-be-seen fluid phases of QCD such as color-flavor locking.\cite{Alford:1997zt} 

There is also a purely theoretical motivation behind this line of research. Symmetries and conserved currents are at the core of quantum field theory. Particle physics has provided us with lessons and examples on a variety of behaviors of symmetry groups: global vs. gauged symmetries, spontaneous symmetry breaking, anomalies etc. On the other hand, thermodynamics, and its extension into hydrodynamics, are essentially theories of the macroscopic evolution of conserved quantities. Thus, hydrodynamics provides a natural framework where the symmetry concepts of QFT can be explored. As a bonus, in the hydrodynamic limit seemingly obscure symmetry-related quantum effects may acquire a macroscopic manifestation. Such is the case with the phenomena of superfluidity and superconductivity, which arise from the spontaneous breaking of a global and gauged symmetry, respectively. 

A fascinating role has been played in this field by the AdS/CFT duality, and by gravitational holography in general. In particular, it is in such a model \cite{Erdmenger:2008rm,Eling:2010hu} that the possibility of axial hydrodynamic transport terms was first noticed. These are contributions to the current known as chiral magnetic and chiral vortical conductivities, and they are now known to take the form:
\begin{eqnarray}
   J_{chiral}^{a\mu} &=& \omega^\mu\left(C^{abc}\mu_b\mu_c + 2\beta^a T^2
     - \frac{2n^a}{h}\left(\frac{1}{3}C^{bcd}\mu_b\mu_c\mu_d + 2\beta^b\mu_b T^2\right)\right) \nonumber \\
     &&+ B_b^\mu\left(C^{abc}\mu_c - \frac{n^a}{h}\left(\frac{1}{2}C^{bcd}\mu_c\mu_d + \beta^b T^2 \right)\right) \ . \label{eq:J_normal}
\end{eqnarray}
The indices $(a,b,\dots)$ label the set of conserved charges. $\omega^\mu = (1/2)\epsilon^{\mu\nu\rho\sigma}u_\nu\del_\rho u_\sigma$ is the vorticity vector of the 4-velocity $u^\mu$, $B_a^\mu$ is the magnetic field, $T$ is the temperature, $\mu_a$ are the chemical potentials, $n^a$ are the charge densities, $h$ is the enthalpy density, while $C_{abc}$ and $\beta_a$ are constant tensors in charge space. $C_{abc}$ is understood \cite{Son:2009tf,Neiman:2010zi,Loganayagam:2011mu} to be the coefficient of the axial $JJJ$ triangle anomaly, while $\beta_a$ is conjectured \cite{Landsteiner:2011cp,Landsteiner:2011iq} to be the coefficient of the gravitational $JTT$ anomaly. Experimental consequences of these effects in heavy-ion collisions have been proposed,\cite{Kharzeev:2010gr,KerenZur:2010zw} and are currently searched for in the collider data.

As we mentioned before, superfluidity is an older example of a macroscopically manifested quantum effect. This extraordinary state of matter has been discovered (in helium)\cite{Kapitza} and described theoretically \cite{Landau,Tisza} already in the early 20th century. Generically, a superfluid contains two coexisting components: a condensate with irrotational velocity characterized by the Goldstone gradient $\xi_\mu$, and a ``normal-fluid'' part characterized by an ordinary 4-velocity $u^\mu$. The early investigations of superfluid helium were of course non-relativistic, and furthermore focused on the limit of vanishing relative velocity between the two components, i.e. $u_\mu \sim \xi_\mu$. Generalizations away from this corner, motivated by more exotic applications, are relatively recent. At the ideal level, relativistic superfluids began to be studied in the early 2000's.\cite{relativistic} The modern study of viscous-order dynamics in relativistic superfluids started from holographic models.\cite{Herzog:2011ec,Bhattacharya:2011ee} This was followed by a general systematic analysis \cite{Bhattacharya:2011tr} of the transport terms allowed by the Second Law of thermodynamics.

\section{The chiral transport terms in superfluids}

Our work is at the intersection of chiral hydrodynamics with superfluidity. Bhattacharya et.al.\cite{Bhattacharya:2011tr} have considered chiral transport terms in superfluids, and came up with a large number of terms consistent with the Second Law. We have generalized this work to the case of additional unbroken charges, and expressed the allowed transport terms in a compact suggestive way. Even this is too cumbersome to reproduce here, and we refer the interested reader to the original paper.\cite{Neiman:2011mj} While in the normal-fluid case all the chiral terms are probably related to anomalies, with superfluids the situation appears to be more delicate. The entropic constraints don't relate any of the new transport terms to anomalies; however, some of the terms have the correct structure to be anomaly-related. We've singled out these terms using analogies to the normal-fluid situation and some heuristic QFT reasoning. This led us to propose a simplified form for the transport coefficients: a quantity that behaves like a $JTT$ anomaly coefficient should be set to a constant, while a quantity that behaves like a $JJT$ anomaly coefficient should vanish, since such an anomaly doesn't exist. With these simplifications, the chiral transport terms consist of the normal-fluid terms (\ref{eq:J_normal}), with the superfluid-specific additions:
\begin{eqnarray}
   T_{chiral-super}^{(1)\mu\nu} &=& \chi^a\pi^{(\mu}_\lambda\epsilon^{\nu)\lambda\rho\sigma}u_\rho\zeta_{a\sigma}
     + a^{abc}\zeta_a^{(\mu}\epsilon^{\nu)\rho\sigma\lambda}u_\rho\zeta_{b\sigma}\pi_{\lambda\kappa}\zeta_c^\kappa \nonumber \\
     &&{}+ b^{abc}_1\zeta_a^{(\mu}\epsilon^{\nu)\rho\sigma\lambda}u_\rho\zeta_{b\sigma}\left(E_{c\lambda} - T\del_\lambda\frac{\mu_c}{T}\right) \label{eq:T_1_short} \\
   J_{chiral-super}^{(1)a\mu} &=& b_2^{abc}\epsilon^{\mu\nu\rho\sigma}u_\nu\zeta_{b\rho}\pi_{\sigma\lambda}\zeta_c^\lambda 
     + c^{abc}\epsilon^{\mu\nu\rho\sigma}u_\nu\zeta_{b\rho}\left(E_{c\lambda} - T\del_\lambda\frac{\mu_c}{T}\right) \label{eq:J_1_short} \\
   \nu_{chiral-super}^{(1)a} &=& \frac{2}{h}\zeta^a_\mu\omega^\mu\left(\frac{1}{3}C^{bcd}\mu_b\mu_c\mu_d + 2\beta^b\mu_b T^2 \right)
      + \frac{1}{h}\zeta^a_\mu B_b^\mu\left(\frac{1}{2}C^{bcd}\mu_c\mu_d + \beta^b T^2\right) \ . \label{eq:nu_1_short}
\end{eqnarray}
Eq. (\ref{eq:T_1_short}) gives the relevant transport contributions to the stress-energy tensor; eq. (\ref{eq:J_1_short}) gives the contributions to the charge current; eq. (\ref{eq:nu_1_short}) gives the correction $\nu^{(1)a} = u^\mu\xi^a_\mu - \mu^a$ to the Josephson relation. These expressions are written in the transverse frame, defined by $T_\mu^{(1)\nu}u_\nu = J_a^{(1)\nu}u_\nu = 0$. $\zeta^a_\mu$ is the component of $\xi^a_\mu$ orthogonal to $u^a_\mu$, i.e. it is proportional to the relative velocity between the two parts of the superfluid. Thus, all the above terms vanish in the limit where the two velocities coincide. We include a charge index on $\xi^a_\mu$ and $\zeta^a_\mu$ in order to clarify the charge structure of the expression, and to suggest its natural generalization to the case of multiple broken symmetry generators. We stress, however, that with multiple broken symmetries new terms are expected to appear, proportional to e.g. $\epsilon^{\mu\nu\rho\sigma}u_\nu\zeta^a_\rho\zeta^b_\sigma$. $\pi_{\mu\nu}$ is the shear tensor associated with the normal-fluid velocity $u^\mu$. $E^a_\mu$ is the electric field. $\chi^a$, $a^{abc}$, $b_1^{abc}$, $b_2^{abc}$ and $c^{abc}$ are transport coefficients. 

Of the above, $\chi^a$, $a^{abc}$ and $b_2^{abc}$ appear clearly unrelated to anomalies: these terms involve the symmetrized velocity gradient $\pi_{\mu\nu}$, while anomalies typically involve curvatures, which are antisymmetrized derivatives. If the physics is invariant under time reversal, $b_1^{abc}$ is tied to $b_2^{abc}$ by the Onsager relation $b_1^{abc} = -b_2^{cba}$. This leads to the conclusion that $b_1^{abc}$ is also unrelated to anomalies. Such non-anomalous chiral transport terms are unique to superfluids, and are speculated to be relevant for condensed-matter systems.\cite{Bhattacharya:2011tr} We are left with $c_{abc}$ as the candidate for a new anomaly-related transport term. This is the topic of the next section.

\section{Chiral electric effect}

Consider the $c_{abc}$ term in (\ref{eq:J_1_short}). It contains a contribution of the form:
\begin{eqnarray}
  J_{CEE}^{a\mu} = c^a{}_{bc}\epsilon^{\mu\nu\rho\sigma}u_\nu\zeta_\rho^b E^c_\sigma \ . \label{eq:chiralE}
\end{eqnarray}
Comparing to the standard electric conductivity term $J_{E}^{a\mu} = \sigma^{ab}E_b^\mu$ and to the anomalous chiral magnetic conductivity $J_{B}^{a\mu} = \sigma_B^{ab}B_b^\mu$, the transport term (\ref{eq:chiralE}) may be dubbed as ``chiral electric conductivity''. The effect consists of an induced current in perpendicular to the applied electric field and to the velocities of both the normal and superfluid components. For this effect to be observable in the bulk of the superfluid, we must have an \emph{unbroken} gauged symmetry in addition to the (global or gauged) broken symmetry characterizing the superfluid. Indeed, even if the broken symmetry is gauged, and thus kinematically capable of developing a field strength $E_\mu$, such a field will be dynamically excluded from the bulk as a result of superconductivity. Superfluid phases with an unbroken gauge symmetry are quite conceivable. A theoretically well-studied example is the color-flavor locking phase of QCD. In this phase, a gauged $U(1)$ generator survives unbroken, and the system behaves with respect to it as an insulator.\cite{Alford:2007xm} The absence of conventional electric conductivity may help to promote the chiral electric conductivity into a dominant effect.

What can be said of the value of the chiral electric coefficient $c_{abc}$? The Second Law of thermodynamics places no restrictions on its value (though it imposes certain inequalities which couple $c_{abc}$ to the ordinary electric conductivity). Given time-reversal invariance, the Onsager principle forces $c_{abc} = c_{cba}$ (and renders the aforementioned inequalities irrelevant). Now, notice that the transport term (\ref{eq:chiralE}) contains only the fluid velocity, gauge field strengths and the Goldstone gradient $\xi^a_\mu$, which is much like a gauge potential. It is therefore tempting to go further and to interpret the coefficient $c_{abc}$ in terms of the axial $JJJ$ anomaly. In our paper,\cite{Neiman:2011mj} we construct an explicit proposal along these lines, motivated by structural analysis of the various transport terms and some heuristic thermal-QFT reasoning. According to this conjecture, the chiral electric conductivity is given by:
\begin{eqnarray}
  c^{abc} = C^{dbe}\left(\delta^a_d - \frac{n^a\mu_d}{h}\right)\left(\delta^c_e - \frac{n^c\mu_e}{h}\right) \ , \label{eq:c}
\end{eqnarray}
where $C_{abc}$ is the same axial anomaly coefficient which appears in the vortical and magnetic conductivities (\ref{eq:J_normal}).

\section{Discussion} \label{sec:discuss}

There are several open issues for future work. Both our simplification (\ref{eq:T_1_short})-(\ref{eq:nu_1_short}) of the transport terms and our proposal (\ref{eq:c}) for the chiral electric conductivity should be tested with a microscopic calculation. Such a calculation with just one (broken) charge will already be a useful check. Also, it will be interesting to have explicit calculations, either thermodynamical or microscopic, for the transport terms in a superfluid with several broken charges. We expect this more general case to be relevant for nuclear and subnuclear fluids, where there are multiple potentially broken generators for the color and flavor symmetries.

The observational relevance of our results, and indeed of previous results along these lines, should be considered. As we mentioned, the transport terms which aren't related to anomalies may have manifestations in nonrelativistic condensed-matter systems. Perhaps there is such hope for the anomalous terms as well - though the anomaly is a relativistic effect, so is magnetism; nonrelativistic velocities do not necessarily preclude the observation of such phenomena. Finally, the color-flavor locked phase of QCD can be further explored as an arena for the chiral electric effect. 

\section*{Acknowledgments}
The work is supported in part by the Israeli Science Foundation center of excellence, by the US-Israel Binational Science Foundation (BSF), and by the German-Israeli Foundation (GIF).

\end{document}